\documentclass[prl,twocolumn]{revtex4}
\usepackage{graphicx}
\renewcommand{\v}[1]{{\bf #1}}

\def\eqa{\begin{eqnarray}}
\def\eea{\end{eqnarray}}
\newcommand{\eq}{\begin{equation}}
\newcommand{\ee}{\end{equation}}

\renewcommand{\>}{\rangle}

\renewcommand{\Re}{{\rm Re}}

\newcommand{\ra}{\rightarrow}

\newcommand{\del}{\delta}
\newcommand{\Del}{\Delta}

\begin{document}

\title{Exact solution of a model of qubit decoherence due to telegraph noise}
\author{Bin Cheng$^1$, Qiang-Hua Wang$^1$ and Robert Joynt$^2$}
\affiliation{$^1$National Laboratory of Solid State Microstructures and Department of
Physics, Nanjing University, Nanjing 210093, China}
\affiliation{$^2$Department of Physics, University of Wisconsin-Madison, Wisconsin 53706,
USA}

\begin{abstract}
We present a general and exact formalism for finding the evolution
of a quantum system subject to external telegraph noise. The
various qubit decoherence rates are determined by the eigenvalues
of a transfer matrix. The formalism can be applied to a qubit
subject to an arbitrary combination of dephasing and relaxational
telegraph noise, in contrast to existing non-perturbative methods
that treat only one or the other of these limits. As the
applications: 1) We obtain the full qubit dynamics on time scales
short compared with the enviromenal correlation times. In the
strong coupling cases this reveals unexpected oscillations and
induced magnetization components; 2) We find in strong coupling
case strong violations of the widely used relation 1/T$_2$ =
1/2T$_1$ + 1/T$_{\phi}$, which is a result of perturbation theory;
3) We discuss the effects of bang-bang and spin-echo controls of
the qubit dynamics in general settings of the telegraph noises.
%The result shows that these methods are not very effective in
%reducing decoherence arising from a single telegraph noise.
Finally, we discuss the extension of the method to the cases of
many telegraph noise sources and multiple qubits. The method still
works when white noise is also present.
\end{abstract}

\maketitle

%\date{\today}

%\date{\today}

\section{I. Introduction}
Decoherence of quantum systems is a fundamental issue with
implications across all branches of physics. In this context,
two-level quantum systems have served as a very useful paradigm.
The subject of two-level systems in random time-dependent fields
originated decades ago in the context of spin resonance
\cite{klauder}. Later experiments on macroscopic
quantum coherence also focused attention on this problem \cite
{leggett}. The recent surge of interest in quantum computing and
quantum control has rejuvenated the field, and much work has gone
into solving various models of decoherence. All physical
realizations of qubits are subject to external noise since all
couple, however weakly, to the external environment. The most
popular models are those that explicitly involve a bath whose
degrees of freedom must be traced out\cite{weiss}. In many
cases, however, one can neglect the flow of quantum information from the
bath to the system, and then it is sufficient to consider the
qubit as being subject to random classical external fields. The
conditions under which this assumption is valid have been
considered in detail by Galperin \textit{et al}. \cite{galperin}.
Here we merely note that when the coupling of the bath to its
thermalizing external environment is very strong or on time scales longer than
the characteristic microscopic times of the bath, we expect that
even fully quantum system-bath models reduce to this case.

The most important type of noise in solid-state systems is
telegraph noise: the qubit interacts with one or more random
fluctuators in its neighborhood, and these fluctuators go back and
forth between only two states \cite{kogan}. Usually the qubit
interacts with many such fluctuators. Depending on the
distribution of fluctuator transition rates this may give rise to
$1/f$ or other types of noise. In this paper, however, we shall
focus on the case of a single fluctuator, and will only mention
the generalization to many fluctuators at the end. This special
case is also extremely important in experimental practice
\cite{single}. Thus we investigate the following model: in a
sequence of $N$ time intervals each of length $\Delta t$, the
fluctuator producing the
noise undergoes a sequence of states labelled by $\left\{
s_{1},s_{2},...,s_{N}\right\} $, $s_{i}=\pm 1$ as shown in Fig. 1, and
so the hamiltonian within the $i$-th interval is given by
\begin{eqnarray}
H_{s_{i}}=-\frac{1}{2}(\v B_{0}+s_i\v g)\cdot\vec{\sigma},
\end{eqnarray}
where $\sigma _{x,y,z}$ are the Pauli matrices acting on the qubit
spin, $\v B_{0}=B_{0}\hat{z}$ pointing in the $z$-direction for
definiteness, and $\pm \v g$ is the effective noise field arising
from the coupling to the fluctuator. The average energy level
separation of the qubit is $B_{0}$ in units where $\hbar =1$ and
$g\mu _{B}=1$. Our notation is that of a spin qubit but the model
obviously applies to any two-level quantum system.

If $\v g$ is in the $\hat z$-direction, then we refer to dephasing
noise, and if $\v g$ is in the x-y plane, then we have relaxational
noise.

\begin{figure}[tbp]
\includegraphics[width=8cm,height=4cm]{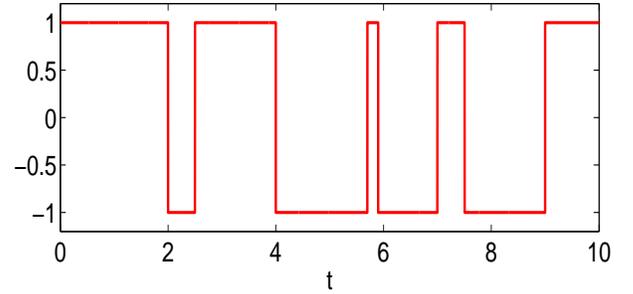}
\caption{One realization of telegraph noise drawn from a distribution with
$p=0.1$, $\del=0$ and $\Del t=0.1$ (in arbitrary units). See the
text for details.}
\end{figure}

We characterize the fluctuator as follows. Within a small time
interval $\Delta t$ the fluctuator state $s^{\prime }$ stays
unchanged, but it changes to $s$ in going to the next interval
with the conditional probability $W_{ss^{\prime }}$. The switching
probabilities are parameterized as $W_{-+}=p+\delta $ and
$W_{+-}=p-\delta$, and normalization requires that
$W_{++}=1-p-\delta$ and $W_{--}=1-p+\delta$. To mimic telegraph
noise we require $p=\gamma \Delta t\ll 1$ and $|\delta |=|\eta|
\Delta t\ll 1$ with finite rates $\gamma $ and $\eta$. For a
fluctuator in thermal equilibrium the stationary level population
of the two fluctuator states is easily shown to be
$p_{s}=(p-s\delta )/2p=(\gamma-s\eta)/2\gamma$ by a standard
detailed-balance argument \cite{machlup}.

A sequence of the noise simulated the model is shown in Fig.1.
We emphasize that telegraph noise has a finite correlation time,
unlike white noise, for which exact methods are already available.
Indeed, the telegraph sequence is a Poisson process and the
noise spectrum is Lorentzian.

We wish to solve for the qubit density matrix $\rho $: given $\rho (t=0),$ find $%
\rho $ at all later times. It has been shown that this decoherence
problem is exactly solvable by methods coming from the theory of
stochastic differential equations.\cite{brissaud} The formal
solution is given in terms of a complicated Laplace transform,
which would be difficult to invert analytically and numerically
unstable in most cases. In special cases the solution can be
simplified and results have been presented when (1)
$g_{x}=g_{y}=0$: pure dephasing ($T_2$) noise
\cite{galperin,paladino,itakura,faoro} (and references therein)
and (2) $g_{z}=0$: pure transverse or relaxational ($T_1$) noise
\cite{itakura,faoro}. These special cases are important. However,
there are clearly many situations in which both types of noise are
present. For example, it has recently been shown that the
character of the noise in flux qubits can be changed continuously
from one type of noise to the other by changing the bias voltage
\cite{kakuyanagi}. The crossover region can only be described by
the more general model, and the only available treatment is to use
the phase-memory functional in the linear-coupling regime where
$g\ll B_{0}$ \cite{bergli}. In this paper, we show how to solve the
general problem for a single qubit by a new algebraic
method. This greatly simplifies the special cases and makes
possible the presentation of results for all values of $\left\{
g_{x},g_{y},g_{z}\right\} $. The solution is exact and does not
make perturbative approximations. The main results are
Eqs.(\ref{Gamma})-(\ref{discreteT}) for the discrete-time
formalism, Eqs.(\ref{P})-(\ref{continuousT}) for the
continuous-time formalism, and the discussions thereafter.

The remainder of this paper is organized as follows. We develop
the exact solution of the problem by the generalized transfer
matrix method in section II, apply the theory to the case of
bang-bang control in section III, and to the case of echo decay in
section IV. We discuss in Sec.V possible generalization of the
theory to the cases of combined white and telegraph noise, many
fluctuators, and many qubits. Finally, Sec.VI is a summary of the
work.

\section{II. transfer matrix ensemble averaged over telegraph noise sequences}

Our task is to solve for $\rho \left( t=N\Delta t\right) $ in the
presence of a time-dependent Hamiltonian, averaged over all
$2^{N}$ sequences with the appropriate probabilities. Formally
\[
\rho (t)=\overline{U_{s_{N}}U_{s_{N-1}}\cdot \cdot \cdot U_{s_{1}}\rho
(0)U_{s_{1}}^{\dag }\cdot \cdot \cdot U_{s_{N-1}}^{\dag }U_{s_{N}}^{\dag }},
\]
where $U_{s}=\exp \left( -iH_{s}\Delta t\right) $ is the evolution
operator at noise level $s$ for one time interval and the overbar
indicates the averaging over the sequences. However, it is much
more convenient to parameterize $\rho$ by
\[
\rho (t)=\frac{1}{2}I+\frac{1}{2}\v n(t)\cdot \vec{\sigma}.
\]
The Bloch vector $\v n(t)$ takes on only real values and
satisfies $|\v n(t)|\leq 1$ (the
equality holds for a pure state). $\ I$ is the $2\times 2$ unit
matrix. Note that $\v n(t)$ also gives the qubit magnetization as
$\langle \vec{\sigma}\>=\mathrm{Tr}(\rho \vec{\sigma})=\v n(t)$.
The time development is given by \eqa \v n\left( t\right) \equiv
T\v n(0)=\overline{T_{s_{N}}T_{s_{N-1}} \cdot \cdot \cdot
T_{s_{1}}}~\v n(0), \label{definetmatrix} \eea which defines the
ensemble averaged transfer matrix $T$, and the matrix $ T_{s}$ is
given by
\[
T_{s}=\exp \left[ i\Delta t\left( B_{0}L_{z}+s\v g\cdot \v L
\right) \right] ,
\]
where $L_{x,y,z}$ are the usual generators of $SO(3)$: $\left(
L_{i}\right) _{jk}=i\epsilon _{ijk}$, where $\epsilon _{ijk}$ is
the completely antisymmetric symbol. The $T_{s}$ are easily
calculated explicitly for arbitrary $\v g$. For noise that is
uncorrelated between time intervals of equal length, the ensemble average in
Eq.(\ref{definetmatrix}) can be performed within each interval,
allowing exact solution of certain noise models \cite{diu}. For true
telegraph noise the ensemble average has to be done in another way which we now
describe.

Let us define $G_{N}^{ss^{\prime }}$ to be the $3\times 3$
transfer matrix for an $N$-step qubit evolution that starts at the
fluctuator state $s^{\prime }$ and end up at the state $s$, but
ensemble averaged over all intermediate fluctuator states. For
$N=1$ no intermediate intervals are involved, so that
$G_{1}^{ss^{\prime }}\equiv \Gamma ^{ss^{\prime }}=W_{ss^{\prime
}}T_{s^{\prime }}$. Here $T_{s^{\prime }}$ is the transfer matrix
in the starting state $s^{\prime }$, and $W_{ss^{\prime }}$
signifies the conditional probability for the change to $s$
immediately after the interval ends. By definition, we find that
\[
G_{N}^{ss^{\prime }}=\sum_{s^{\prime \prime }}\Gamma ^{ss^{\prime \prime
}}G_{N-1}^{s^{\prime \prime }s^{\prime }},
\]%
which is already in the form of a matrix product, and by iteration we see
that $G_{N}=\Gamma ^{N}$, where we defined an operator
\begin{eqnarray}
\Gamma  &=&(1-p-\delta \tau _{3}+p\tau _{1}-i\delta \tau _{2})\times
\nonumber \\
&&\exp (i\Delta t\v B_{0}\cdot \v L+i\Delta t\v g\cdot \v L\tau
_{3}),\label{Gamma}
\end{eqnarray}%
where the Pauli matrices $\{\tau _{i}\}$ act on the fluctuator
state $|s\>$ (the eigenstate of $\tau _{3}$). It is easily seen
that $\Gamma ^{ss^{\prime }}=\langle s|\Gamma |s^{\prime }\>$.
Finally the globally ensemble-averaged transfer matrix is given by
\eqa T=\sum_{s,s^{\prime }}G_{N}^{ss^{\prime }}p_{s^{\prime
}}=\langle x_{f}|\Gamma ^{N}|i_{f}\>, \label{discreteT}\eea where
$|x_{f}\>=\frac{1}{\sqrt{2}}\sum_{s=\pm }|s\>$ is formally one of
the eigenstate of $\tau _{1}$, and $|i_{f}\>=\sqrt{2}\sum_{s=\pm
}p_{s}|s\>$ encodes the initial stationary level distribution of
the fluctator. Note that the formal inner product is performed in
the fluctuator level space, leaving a $3\times 3$ matrix acting on
the initial qubit vector $\v n(0) $.

In the limit of $\Delta t\rightarrow 0$,
\begin{eqnarray}
\Gamma && \rightarrow 1+i\Delta t (\v B_0L_z +\v g\cdot \v L\tau_3)
\nonumber \\
&& -\Delta t \gamma+\Delta t( \gamma\tau_1-i\eta\tau_2-\eta\tau_3)  \nonumber
\\
&& \sim \exp\left(-\Delta t P\right),
\end{eqnarray}
where we define
\begin{eqnarray}
P&&=\gamma-i\v B_0L_z-i\v g\cdot\v
L\tau_3-\gamma\tau_1+i\eta\tau_2+\eta\tau_3.\label{P}
\end{eqnarray}
Thus in the continuum time limit $G=\Gamma^N=\exp(-tP)$ and so
\begin{eqnarray}
T=\langle x_f|\exp(-tP)|i_f\> \label{continuousT}
\end{eqnarray}
for $t=N\Delta t$, and the problem reduces to the diagonalization of $P$
which can be cast into a $6\times 6$ matrix. Assuming that $P$ is not
defective (an assumption we have checked in the cases treated here), $T$ can
be decomposed as
\begin{eqnarray}
T=\sum_\lambda \langle x_f|\lambda\>\langle\lambda|i_f\> \exp(-\lambda t),
\end{eqnarray}
where $|\lambda\>$ and $\langle\lambda|$ are the right and left eigen
vectors of $P$ with the eigenvalue $\lambda$, normalized such that $%
\langle\lambda|\lambda^{\prime}\>=\delta_{\lambda\lambda^{\prime}}$.
(Notice again the partial inner products with $|i_f\>$ and
$\langle x_f|$.) Each eigenvalue corresponds to a relaxation time
of the system. Typically, the shorter times correspond to
transients and the longest two times correspond to $T_{1}$ and
$T_{2}$ - this can be verified by examining $\v n(t))$ in detail.

%Usually, it is more convenient to analyze the qubit evolution in
%the rotating frame, where the qubit vector is given by $\tilde{\v
%n}(t)=R^{-1}\v n(t)$ with $R=\exp(itB_0L_z)$. The transfer matrix
%in such a frame is therefore given by $\tilde{T}=R^{-1}T$. The
%advantage of using rotating frame is that intrinsic oscillations
%at frequencies $\pm B_0$ are filtered away.

Now let $\theta $ be the angle between $\v g$ and the $z-$axis. For $%
\theta =0$, $\v g=g\hat{z}$, $L_{z}$ is conserved by $P$, so that
we can use the quantum numbers $m={0,\pm 1}$ in place of $L_{z}$,
and diagonalize the $2\times 2$ matrix in the fluctuator spin
space. The eigenvalues $\lambda $ are given by
\[
\lambda =\gamma -iB_{0}m\pm \sqrt{\gamma ^{2}-g^{2}m^{2}-2ig\eta m},\ \ \
m=0,\pm 1.
\]%
The eigenvectors, and finally the transfer matrix $T$ can also be easily
obtained. We shall not go into these details here. By inspection of $\Re
(\lambda )$, we see that the relaxation rate $1/T_{1}=0$ in the $n_{z}(t)$
channel ($m=0$), whereas the dephasing rate $1/T_{2}=\gamma -\Re \sqrt{%
\gamma ^{2}-g^{2}\pm 2ig\eta }$ in the $n_{x,y}$ channel. The result agrees
with the results in Ref. \cite{paladino}.

The other case for which compact explicit expressions can be give is
$\theta =\pi/2$ (so that $\v g$ is perpendicular to $\v B_{0}$) and at the
same time the switching rate imbalance $\eta =0$. For simplicity
let us assume that $\v g$ is in x-direction and $\v B_{0}$
is still in the z-direction. Upon a rotation in the level space,
we have $U\tau _{1}U^{\dagger }=\tau _{3}$ and $U\tau
_{3}U^{\dagger }=-\tau _{1}$, where $U $ is the SU(2) rotation
about the y-axis by 90 degrees (in the level spin space). Under
this transformation
\[
UPU^{\dagger }=\gamma -\gamma \tau _{3}-iB_{0}L_{z}+igL_{x}\tau _{1}.
\]%
This matrix is the same as that of a Hamiltonian for the
coupling of a spin-1 particle with angular momentum $\v L$ and
a spin-1/2 particle with angular momentum $\v S=\vec{\tau}/2$.
Inspection reveals that the operator only mixes states whose
z-components of $\v L+\v S$ differ by 2. (This is seen by writing $L_{x}=(L_{+}+L_{-})/2$ and $%
S_{x}=\tau _{1}/2=(S^{+}+S^{-})/2$). As the result, the Hilbert space is
divided into two invariant subspaces, spanned respectively by the two sets
of states
\begin{eqnarray}
&&\{|-1,-1/2\>,|1,-1/2\>,|0,1/2\>\},  \nonumber \\
&&\{|-1,1/2\>,|0,-1/2\>,|1,1/2\>\}.  \nonumber
\end{eqnarray}%
Here in each basis state the first index refers to $L_{z}$, and
the second one to that of $S_{z}=\tau _{3}/2$. The diagonalization
can carried out separately in the two subspaces, and only $3\times
3$ matrices are involved. The T-matrix is formally given by
$T=\langle z_{f}|\exp (-tUPU^{\dagger })|z_{f}\>$, where
$|z_f\>=(1,0)^T$. Here we used the facts that $|i_{f}\>=|x_{f}\>$
for $\eta =0$ and that $U|x_{f}\>=|z_{f}\>$. Without going into
details we mention that the eigenvalues of $UPU^{\dagger }$ (and
thus of $P$) satisfy one of the two equations below:
\begin{eqnarray}
&&\lambda ^{3}+2\gamma \lambda ^{2}+(B_{0}^{2}+g^{2})\lambda
+2B_{0}^{2}\gamma =0, \\
&&\lambda ^{3}+4\gamma \lambda ^{2}+(B_{0}^{2}+g^{2}+4\gamma ^{2})\lambda
+2g^{2}\gamma =0.
\end{eqnarray}%
The result can also be obtained from the stochastic differential equation
approach \cite{itakura}.

\begin{figure}[tbp]
\includegraphics[width=8cm]{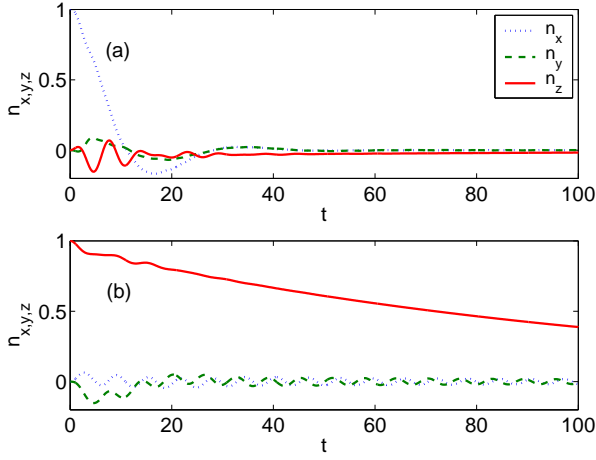}
\caption{(Color online) The evolution of the qubit vector $\v n=(n_x,n_y,n_z)
$ in the rotating frame, starting from (a) $\v n(0)=\hat{x}$, and (b) $\v
n(0)=\hat{z} $. The parameters are $B_0=1$, $\protect\gamma=0.1$, $g=0.3$, $%
\protect\eta=0 $, and $\protect\theta=\protect\pi/4$. The time $t$ is in
units of $1/B_0$. The same legend for the curves is used in (a) and (b).}
\end{figure}

\begin{figure}[tbp]
\includegraphics[width=8cm]{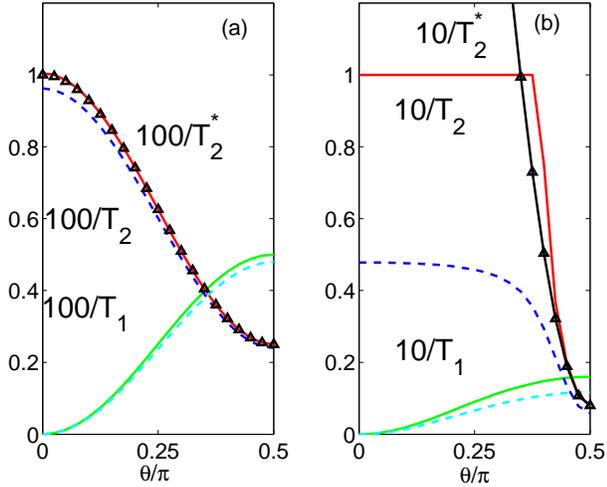}
\caption{(Color online) The relaxation rate $1/T_{1}$ and dephasing rate $%
1/T_{2}$ (in units of $B_{0})$ as functions of the angle $\protect\theta $
between the static field $\v B_{0}$ and the noise field $\v g$. $1/T_{2}^{*}$
(triangle points) is $1/T_2$ calculated from the perturbative
formula in Eq. 11.  The weak coupling case (a) has the parameters
$\protect\gamma =0.5$ and $g=0.1$. Here $\protect\eta =0$ ($%
0.1$) for solid (dashed) lines), while the strong coupling case (b) uses $\protect%
\gamma =0.1$ and $g=0.3$. Here $\protect\eta =0$ ($0.05$) for solid (dashed)
lines. Rescaling of the rates is used to improve clarity. The perturbative formula
breaks down completely for strong coupling. }
\end{figure}

For general $\theta $, the analysis (in terms of phase memory function) in
the literature is limited so far to the so-called linear-coupling regime $%
g\ll B_0$ \cite{bergli}. Our model is however exact for any coupling
strength. All that we have to do is to diagonalize a $6\times 6$ matrix to
obtain the full qubit dynamics. In Fig.2 we plot the evolution of the qubit
vector $\v n$ in the rotating frame defined by $\v B_{0}$. The parameters
are $B_{0}=1$, $\gamma =0.1$, $g=0.3$, $\eta =0$ and $\theta =\pi /4$. We
see from Fig.2(a) that starting from $\v n (0)=\hat{x}$, $n_x$ decays with
oscillations even in the rotating frame. On the other hand the $z$-component
is induced in the intermediate stage and actually decays more slowly than
the $x$-component. We checked that this feature is visible in the
strong coupling regime ($\gamma< g$), but is much weaker in the weak
coupling regime ($\gamma> g$), and is completely absent in the case of $%
\gamma\gg g$, the limit of white noise. This signifies the unique role of long
time correlations. In Fig.2(b) the qubit starts from $\v n=\hat{z}$ and
decays slower than the x- and y-components induced in the intermediate
stages, in agreement with the behavior in Fig.2(a).

One should be aware that none of this qubit dynamics is available from
Redfield theory, which applies only to the regime $\gamma t \gg 1$.

We can decide precisely the asymptotic decay rates in the different
channels by matching $n_{x,y,z}(t)$ with the envelope curves $\exp
\left[ -\Re \left( {\lambda }\right) t\right]$, where $\lambda$ are
the numerical eigenvalues of $P$. The resulting decay rates
$1/T_{1}$ (in the $n_{z}$-channel) and $1/T_{2}$ (in the $n_{x,y}
$-channel) as a function of $\theta $ are plotted in Fig.3. First
consider the case of $\eta =0$ (solid lines). Fig.3(a) is in the
weak-coupling regime, where we see a smooth change and a crossing of
the two rates as $\theta $ increases. This is similar to the case of
uncorrelated noise\cite{tahan}. The reason is that for $g<\gamma$,
in a time scale determined by $1/g$ many switches occur and as such
the noise is essentially uncorrelated beyond a time scale of $1/g$.
Fig.3(b) is in the strong-coupling regime, where we note the
transition from the flat behavior of $T_{2}$ at small angles to a
downturn at larger angles. This is because $1/T_{2}$ is largely
controlled by $g_{z}$, which decreases with increasing $\theta $ and
eventually falls into an "effective weak-coupling" regime $
g_{z}<\gamma $ for $1/T_2$. We also observe that $T_{2}=2T_{1}$ at
$\theta =\pi /2$ for both $g>\gamma $ and $g\leq \gamma $, and is
therefore a consequence of the model for all coupling strengths
(provided that $\eta =0$). This can be checked analytically from the
eigenvalue equations in the particular limits of $\gamma\rightarrow
0$ or $g\rightarrow 0$. Second, we consider the effect of a nonzero
$\eta $ (dashed lines). As compared to the $\eta =0$ case, we see
that in the strong coupling case (b) the flat regime and the sharp
transition of $1/T_{2}$ are smeared, and in both Fig.3(a) and
Fig.3(b) the decay rates become smaller. This is understandable from
the fact that a nonzero $\eta $ amounts to a nonzero average of the
noise field $\v g$ and thus a decrease of the amount of fluctuating
component. In particular, we have checked that $1/T_{1}=1/T_{2}=0$
for $\eta =\pm \gamma $, as one would have anticipated since the
fluctuator stops at one of the two levels and does not switch at
all.

In the literature the decay rates at arbitrary $\theta$ are
available only in the weak coupling cases. The perturbative
results are often summarized as \cite{slichter} \eqa 1/T_2 =
1/2T_1+1/T_{\phi}\eea with \eqa 1/T_{\phi}=\cos^2\theta S(0)/2,\\
          1/T_1=\sin^2\theta S(\Omega)/2.\eea
Here $1/T_{\phi}$ is the dephasing rate as if in an effective
z-direction random field with amplitude $B_0\cos\theta$ alone,
$S(\omega)$ is the power spectrum of telegraph noise and
$\Omega=B_0$. In view of the wide use of this formula, it is
important to check how well it holds for general coupling
strengths. The comparison is made in Fig.3 (for $\eta=0$ only),
where $1/T_2$ calculated according to the above formula is denoted
by $1/T_{2}^{*}$ (triangles) and thus can be compared to the
actual $1/T_2$. We see that in the weak (strong) coupling case of
Fig.3(a) [Fig.3(b)] it agrees perfectly with (deviates
considerably from) our exact result (solid lines). The formula
cannot be used when the coupling of the source to the qubit is
large compared to the inverse correlation time of the noise. This
illustrates the importance of our exact results in the strong
coupling cases.

\section{III. dynamical decoupling by bang-bang control}

The formalism can be easily adapted to echo decay measurements and
the bang-bang control protocol where control pulses are applied. In
such processes the transfer matrix can be formally written as
\[
T=\langle x_{f}|\hat{T}\exp [-\int_{0}^{t}dt^{\prime }P+i\sum_{i}\vec{\phi}%
(t_{i})\cdot \v L]|i_{f}\>,
\]%
where $\hat{T}$ time-orders the operators, encoding the
instantaneous rotations with vectors $\{\vec{\phi}(t_{i})\}$
caused by pulses at time $\{t_{i}\}$. These manipulations are
important for comparison with experiments such as that in
Ref.\cite{kakuyanagi}.

Let us consider an open loop quantum control (or bang-bang
control), in which a sequences of $\pi_{x}$ or $\pi_{y}$ pulses
with fixed intervals are applied to reduce the decoherence due to
slowly low frequency noise. In the ideal case, we assume that each
pulse is of zero width in time. The problem of ideal dynamical
decoupling by bang-bang control has been exactly solved in the
special cases of pure dephasing noise ($\theta=0$) and pure transverse noise
($\theta = \pi/2$) \cite{faoro}. With our method
we can get the exact solution at arbitrary working points.

In some cases $\pi_{x}$ pulses would cause the anti-Zeno effect in
experiment\cite{kofman}, so that we consider $\pi_{y}$ pulses.
After a $\pi_{y}$ pulse is applied, the Bloch vector is rotated by
$\pi$ about the $y$-axis: $\v n\ra \exp(i\pi L_y) \v n$. For a
periodic sequence of $\pi_y$ pulses, the transfer matrix is
explicitly given by
\begin{eqnarray}
T(N\tau)=\langle x_{f}|\left[\exp(-P\tau)\exp(i\pi
L_y)\right]^N|i_f\rangle,
\end{eqnarray}
where $\tau$ is the interval between two adjacent pulses, $N$ is
the number of pulses applied. By diagonalizing the operator (again
a $6\times6$ matrix) $\exp(-P\tau)\exp(i\pi L_y)$, we get the
eigenvalues $\lambda_{i=1,...,6}$ and the candidate decay rates
$\Gamma_{i}=-\ln|\lambda_{i}|/\tau$. In principle only one of the
six $\Gamma$'s control the long time asymptotic behavior in a
specific channel. This rate can be decided precisely, as we used
in Sec.II, by matching $n_{x, y, z}(t)$ with the envelope curve
$\exp(-\Gamma_i t)$.

\begin{figure}[tbp]
\includegraphics[width=8cm]{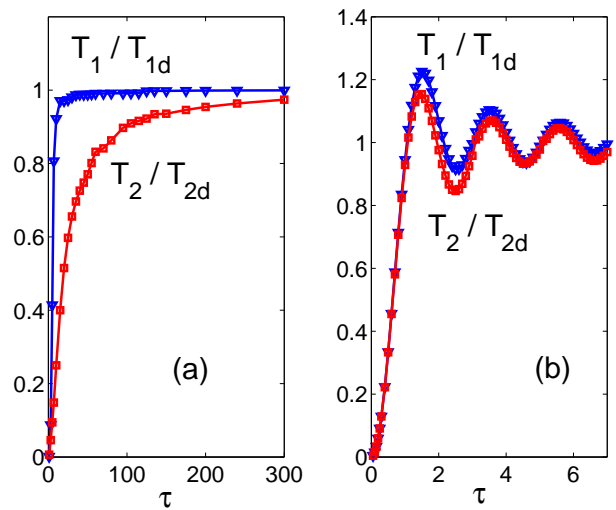}
\caption{(a)The normalized rates for bang-bang control. The
relaxation rate $T_{1}/T_{1d}$(triangles) and the dephasing rate
$T_{2}/T_{2d}$(squares) are plotted as functions of the interval
$\tau$ between $\pi_{y}$ pulses, with the parameters $B_{0}=1$,
$\gamma=0.1$, $g=0.03$, $\eta=0$, and $\theta=\pi/4$. (b) The same
plot as (a) except that $g=3$.  $1/T_{1}$ and $1/T_{2}$ are the
rates in the absence of the pulses.  Lines are drawn to guide the
eye.}
\end{figure}

In Figs.4 we plot the normalized decay rates $T_{1}/T_{1d}$ (in
the $n_{z}$-channel) and $T_{2}/T_{2d}$ (in the $n_{x,y}$-channel)
as functions of the interval $\tau$ between pulses. $1/T_{1}$ and
$1/T_{2}$ are the rates in the absence of the pulses.  A
normalized rate of 1 therefore corresponds to no suppression of
decoherence by the pulses. We set $B_{0}=1$, $\eta=0$,
$\theta=\pi/4$, and $\gamma=0.1$. In addition, $g=0.03$ and $g=3$
in Fig.4(a) and (b), respectively. We observe that as the interval
$\tau$ between $\pi_{y}$ pulses decreases, the decay rates of both
$n_{z}$-channel and $n_{x,y}$-channel decrease. The reduction is
significant as soon as $\tau\sim 1/g$. Moreover, in Fig.4(b) where
$\theta=\pi/4$ we observe oscillatory behavior of the relative
rates similar to the case of $\theta=\pi/2$ studied
elsewhere.\cite{faoro}
%Thus bang-bang-control is not very effective in cancelling out
%telegraph noise.

\section{IV. Spin-echo decay}
In a spin-echo process, a $\pi_x$ pulse is applied half-way
between two $\pi_x/2$ pulses. The first $\pi_x/2$ pulse rotate the
initial quibt state $n_z$ to $n_y$, the second $\pi_x/2$ rotate
$n_y$ back to $n_z$ right before the measurement. The $\pi_x$
pulse reverses the $n_y$-component, reducing the line-broadening
associated with low frequency noise in the final measurement. The
transfer matrix in the combined process is easily shown to be
given by
\begin{eqnarray}
T(t)=&&\langle x_f|\exp(i\pi L_x/2)\exp(-t P/2)\exp(i\pi
L_x)\nonumber\\ && \times \exp(-t P/2)\exp(i\pi
L_x/2)|i_{f}\rangle,
\end{eqnarray}
where $t$ is the time interval between the two $\pi_x/2$ pulses.

\begin{figure}[tbp]
\includegraphics[width=8cm]{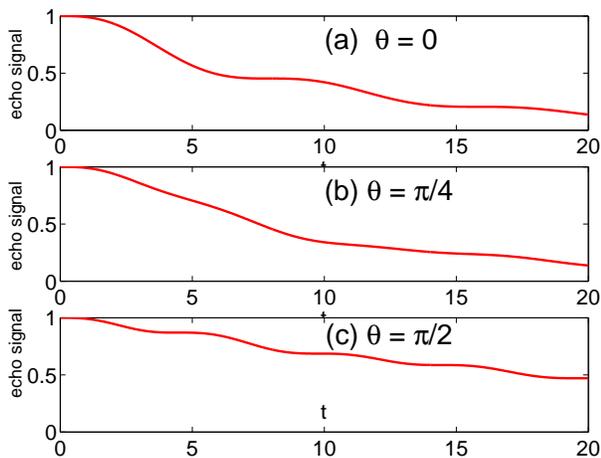}
\caption{The time dependence of echo signals for $g=0.8$,
$\gamma=0.1$ and (a) $\theta=0$(pure dephasing), (b)
$\theta=\pi/4$ and (c) $\theta=\pi/2$. }
\end{figure}

In Fig.5 we plot the time dependence of the echo signal for the
parameters $g=0.8$ (strong coupling case), $\gamma=0.1$, and (a)
$\theta=0$, (b) $\theta=\pi/4$ and (c) $\theta=\pi/2$.  We can see
that the echo signal shows steps. The steps in the special case of
$\theta=0$ have been observed experimentally,\cite{nakamura} as
has been pointed out in Ref.\cite{galperin}. Here we point out
that steps should occur for all $\theta$, and the period of the
steps depends sensitively on $\theta$.  Indeed, the longest period
and the most pronounced steps occur for $\theta = \pi/4$.

\begin{figure}[tbp]
\includegraphics[width=8cm]{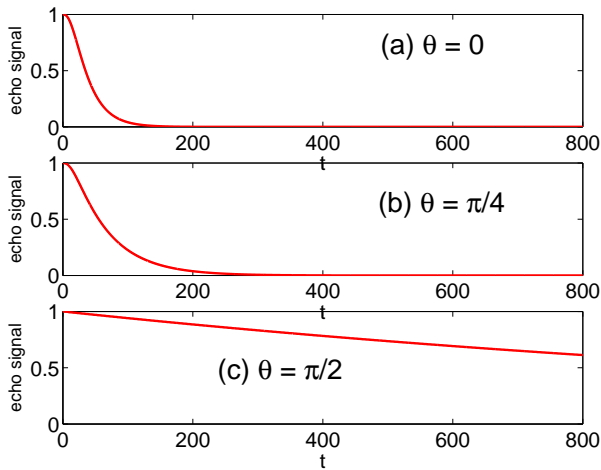}
\caption{The time dependence of echo signal for $g=0.08$ and $\gamma=0.1$.
(a) gives the case $\theta=0$, (b) is the case $\theta=\pi/4$
and (c) is the case $\theta=\pi/2$, which corresponds to the
experiment in Ref. \cite{kakuyanagi}.}
\end{figure}

Figs.6 are the same plots as Figs.5 except that $g=0.08$ (weak
coupling case). In this case no steps are observed. The signal
decays exponentially. This is in agreement with the experiment at
the optimal working point ($\theta = \pi/2$)\cite{kakuyanagi}, but
our results show that this is a general behavior for any values of
$\theta$ and the decay is faster for $\theta<\pi/2$ (which is
reasonable since the z-component of the noise field increases as
$\theta$ decreases). The echo decay rate in the one fluctuator
model is not reduced significantly as compared to free-induction
decay, while in the experiment of Ref. \cite{kakuyanagi,nakamura}
the reduction is very significant. This indicates that a
many-fluctuator model (plus possibly other types of noise) is
needed.  This will be discussed in the next section.

\section{V. Generalizations}
The formalism developed so far can be extended to the case of
combined noise sources: a white-noise field $\v h$ and a
two-level fluctuator. We assume that in a time sequence $\v h$
does not change within an interval $\Delta t$, but is uncorrelated
between different intervals. This enables us to do the ensemble
average over $\v h$ within each time interval, where we assume
$\langle h_i h_j\>=(v_i/\Delta t)\delta_{ij}$, with $i=x,y,z$. The
effect of $\v h$ enters $\Gamma$ as
\begin{eqnarray}
\Gamma=&&(1-p-\delta\tau_3+p\tau_1-i\delta\tau_2)\times  \nonumber \\
&& \exp[i\Delta t(\v B_0+\v h)\cdot \v L+i\Delta t\v g\cdot \v
L\tau_3].
\end{eqnarray}
Expanding the exponential function (up to the second order of $\v
h$) and performing the ensemble average over $\v h$, and finally
taking the limit $\Delta t\rightarrow 0$ (recalling that $p=\gamma
\Delta t$ and $\delta=\eta\Delta t$) we again obtain
$\Gamma=\exp(-\Delta t P)$ with a modified operator
\begin{eqnarray}
P=&&\gamma+\sum_i v_i L_i^2/2 -i\v B_0L_z  \nonumber \\
&& -i\v g\cdot\v L\tau_3-\gamma\tau_1+i\eta\tau_2+\eta\tau_3.
\end{eqnarray}
Here we assumed that the contributions from higher order moments
of $\v h$ are of higher orders in $\Delta t$ and can be ignored.
As such the result is independent of the concrete form of the
distribution function for $\v h$.

The theory can also be generalized to many independent fluctuators
described by the parameters $\{ \v g_n, \gamma_n, \eta_n\}$
($n=1,...,N$). We assume that after each small time interval at
most one of the fluctuators could switch (joint switches occur at
higher orders of $\Delta t$). Since the fluctuators are
independent, it is straightforward to see that the desired $P$
operator is now given by
\begin{eqnarray}
P=&&\sum_n(\gamma_n-i\v g_n\cdot\v
L\tau_{3n}-\gamma_n\tau_{1n}+i\eta_n\tau_{2n}+\eta_n\tau_{3n})\nonumber\\
&&-iB_0L_z,
\end{eqnarray}
where $\tau_{i n}$ denotes the $i$-th Pauli matrix operating on
the $n$-th fluctuator spin. This is a system with a spin-1 coupled
to many independent spin-1/2's. The problem is still solvable
algebraically provided that $ [L_z,P]=0$. In general, $P$ is a
$(3 \times 2^N) \times (3 \times 2^N)$ matrix.
The final $3\times 3$ T-matrix according to which the Bloch vector evolves
is formally given by $T=\langle x_f|\exp(-tP)|i_f\> $ with
$|x_f\>=\Pi_n |x_{fn}\>$ and $|i_f\>=\Pi_n |i_{fn}\>$, where $
|x_{fn}\>$ and $|i_{fn}\>$ describe the states of the $n$-th
fluctuator.

Finally the method can be generalized to the case of many qubits.
Let $D$ be the dimension of the Hilbert space of the qubits. Then
there are $D^{2}-1$ Hermitian generators $X_{i}$ of the
transformation group $SU(D)$ instead of the three Pauli matrices
used in the single qubit case where $D=2$. The $D\times D$ density
matrix may be expanded as a real linear combination of the $X_{i}$
as in Eq.(2):
\[
\rho \left( t\right) =I/D+\sum_{i=1}^{D^{2}-1}a_{i}\left( t\right)
~X_{i}.
\]
The $\left( D^{2}-1\right) \times \left( D^{2}-1\right) $ matrix
$T_{s}$ that evolves the vector $\v a (t) $ forward in time when
the joint state $s$ of the $N$ two-level fluctuators are given is
now easily computed, and the steps in deriving the transfer matrix
are formally identical to that in Sec.II. We see that the
corresponding operator $P$ will be a
$(2^N(D^{2}-1)) \times (2^N(D^{2}-1))$ matrix, and its eigenvalues determine the $\left(
D^{2}-1\right) $ relaxation times in the system, one corresponding
to each possible observable. The rapidly increasing
dimension of the superoperator space will begin to give difficulties
for numerical calculations even at moderate values of $N$ and $D$.

\section{VI. Summary}
We have developed an exact transfer matrix method to solve the
problem of qubit decoherence caused by one fluctuator, and
have applied the theory to free qubit decay, bang-bang control and spin-echo
decay. We have reproduced all known exact results and have shown that
the perturbative limits are correct.  The method is relatively
straightforward to apply, since it is completely algebraic, and the
qubit decoherence rates are determined by the eigenvalues of
a transfer matrix.

The formalism is also more powerful than previous exact methods in
that it can be applied to a qubit subject to an arbitrary
combination of dephasing and relaxational telegraph noise. The
typical way of combinnig the two types of noise is according to
the perturbative formula 1/T$_2$ = 1/2T$_1$ + 1/T$_{\phi}$.  We
have shown that this formula breaks down when the coupling is
strong. The algebraic method can get qubit dynamics on time scales
short compared with the environmental correlation times, which is
often of experimental interest.
%We also find that bang-bang
%coupling and spin-echo methods are not very effective in reducing
%decoherece arising from telegraph noise.

The method can be generalized to the case of many noise sources
and multiple qubits, though the size of the matrices grows rapidly.

\acknowledgments{We would like to thank S. N. Coppersmith and D.
Nghiem for useful conversations. The work in Nanjing was supported
by NSFC 10325416, the Fok Ying Tung Education Foundation No.91009,
the Ministry of Science and Technology of China (under the Grant No.
2006CB921802 and 2006CB601002) and the 111 Project (under the Grant
No. B07026). The work in Wisconsin was supported by NSF-ITR-0325634
and NSF-EMT-0523680}.

\end{document}